\documentstyle[12pt,aaspp4]{article}

\lefthead{Machida et al.}
\righthead{Global 3D Simulations of Magnetized Tori}

\begin{document}

\title{GLOBAL SIMULATIONS OF DIFFERENTIALLY ROTATING MAGNETIZED DISKS:
FORMATION OF LOW-BETA FILAMENTS AND STRUCTURED CORONA}

\author{M. MACHIDA}
\affil{Graduate School of Science and Technology, Chiba University,
1-33 Yayoi-Cho, Inage-ku, Chiba 263-8522, Japan} 

\author{MITSURU R. HAYASHI}
\affil{National Astronomical Observatory, Mitaka, Tokyo 181-8588, Japan}

\and

\author{R. MATSUMOTO}
\affil{Department of Physics, Faculty of Science, Chiba University,
1-33 Yayoi-Cho, Inage-ku, Chiba 263-8522, Japan}

%

\begin{abstract}
We present the results of three-dimensional global magnetohydrodynamic
(MHD) simulations of the Parker-shearing instability in a
differentially rotating torus initially threaded by toroidal magnetic
fields. An equilibrium model of magnetized torus is adopted as an
initial condition. When $\beta_0 = P_{\rm gas}/P_{\rm mag} \sim 1$ at
the initial state, magnetic flux buoyantly escapes from the disk and
creates loop-like structures similar to those in the solar corona.
Inside the torus, growth of non-axisymmetric magneto-rotational (or 
Balbus \& Hawley) instability generates magnetic turbulence. Magnetic
field lines are tangled in small scale but in large scale they show
low azimuthal wave number spiral structure. After several rotation
period, the system oscillates around a state with $\beta \sim
5$. We found that magnetic pressure dominated ($\beta < 1$) filaments
are created in the torus. The volume filling factor of the region
where $\beta \leq 0.3$ is 2-10\%. Magnetic energy release in such
low-$\beta$ regions may lead to violent flaring activities in accretion
disks and in galactic gas disks. 

\end{abstract}


\section{INTRODUCTION}

Magnetic fields in differentially rotating disks play essential roles 
in the angular momentum transport which enable the accretion and
various activities such as X-ray flares and jet formation. Motivated
by the Skylab observations of the solar corona, Galeev, Rosner, \&
Vaiana (1979) proposed a model of magnetically structured corona in
accretion disks consisting of X-ray emitting magnetic loops. The
magnetic loops can be created due to the buoyant rise of magnetic flux
tubes (or flux sheet) from the interior of the accretion
disk. Matsumoto et al. (1988) carried out two-dimensional
magnetohydrodynamic (MHD) simulations of the Parker instability
(Parker 1966) in nonuniform gravitational fields which mimic those in
accretion disks. They showed that when $\beta = P_{\rm gas}/P_{\rm
mag} \sim 1$, a plane-parallel disk deforms itself into evacuated
undulating magnetic loops and dense blobs accumulated in the valley of
magnetic field lines. The effects of rotation and shear flow, however,
were not included in their simulations. 
  
Shibata, Tajima, \& Matsumoto (1990) carried out two-dimensional MHD
simulations of the Parker instability including the effects of shear
flow and suggested that magnetic accretion disks fall into two types;
gas pressure dominated (high-$\beta$) solar type disks, and magnetic
pressure dominated (low-$\beta$) cataclysmic disks. In high-$\beta$
($\beta \geq 1$) disks, magnetic flux escapes from the disk more
efficiently as $\beta$ decreases.

Several authors (Hawley, Gammie \& Balbus 1995 ; Matsumoto \& Tajima
1995 ; Brandenburg et al. 1995; Stone et al. 1996) have reported the
results of three-dimensional local MHD simulations of magnetized
accretion disks by adopting a shearing box approximation (Hawley et
al. 1995). In differentially rotating disks, magnetorotational
instability (Balbus \& Hawley 1991) couples with the Parker
instability (Foglizzo \& Tagger 1995).  
Since Parker instability grows for long wave length perturbations
along magnetic field lines, non-local effects may affect the stability
and nonlinear evolution. 

Results of global 3D MHD simulations including vertical gravity were
reported by Matsumoto (1999) and Hawley (1999). By adopting an
initially gas pressure dominated torus threaded by toroidal magnetic
fields, Matsumoto (1999) showed that magnetic energy is amplified
exponentially owing to the growth of the Balbus \& Hawley instability
and that the system approaches a quasi-steady state with $\beta \sim
10$. Matsumoto (1999) assumed $\beta=constant$ in the torus at the
initial state. When $\beta_0 \sim 1$, the deviation from
magneto-rotational equilibrium introduces large amplitude
perturbations.   

In this paper, we present the results of 3D MHD simulations starting
from an equilibrium MHD torus threaded by initially equipartition
strength ($\beta \sim 1$) toroidal magnetic fields. 


\section{MODELS AND NUMERICAL METHODS}

The basic equations we use are ideal MHD equations in cylindrical 
coordinate system $(r,\phi ,z)$.
We assume that the gas is inviscid and evolves adiabatically. 
Since we neglect radiative cooling, our numerical simulations
postulates that the disk is advection-dominated (see Kato, 
Fukue, \& Mineshige 1998 for a review).

The initial condition is an equilibrium model of an axisymmetric
MHD torus threaded by toroidal magnetic fields (Okada et al. 1989).  
We assume that the torus is embedded in hot, isothermal,  
non-rotating, spherical coronal gas. For gravity, we use the 
Newtonian potential. We neglect the self-gravity 
of the gas. At the initial state the torus is assumed to have a 
constant specific angular momentum $L$. 

We assume polytropic equation of state $P= K \rho^{\gamma}$ 
where $K$ is a constant, and $\gamma$ is the specific heat ratio. 
 According to Okada et al. (1989), we assume
\begin{equation}
v_{A}^2 = \frac{B_{\phi}^2}{4 \pi \rho} =  H ( \rho r^2)^{\gamma-1} 
\end{equation}
where $v_A$ is the Alfv\'en speed and $H$ is a constant. For
normalization we take the radius $r_0$ where the rotation speed
$L/r_0$ equals the Keplerian velocity $v_{K0}=(GM/r_0)^{1/2}$ 
as unit radius and set $\rho_0=v_{K0}=1$ at $r=r_0$.
Using these normalizations, we can integrate the
equation of motion into the potential form;
\begin{equation}
\Psi = -\frac{1}{R} + \frac{L^2}{2 r^2}+\frac{1}{\gamma-1}v_s^2
+\frac{\gamma}{2(\gamma-1)}v_A^2 = \Psi_0 = constant ,
\end{equation}
where $v_s^2$ is the square of the sound speed, 
$R = (r^2+z^2)^{1/2}$ and $\Psi_0=\Psi(r_0,0)$.
By using equation (2), we obtain the density distribution.

\begin{equation}
\rho= \left\{
\frac{{\rm max}[\Psi_0 + 1/R -L^2/(2 r^2) , 0]}
{K[\gamma/(\gamma-1)][1+\beta_0^{-1} r^{2(\gamma-1)}]}
\right\} ^{1/(\gamma-1)}
\end{equation}
where $\beta_0 \equiv 2K/H$ is the plasma $\beta$ at $(r,z)=(r_0,0)$. 
The parameters describing the structure of the MHD torus are 
$\gamma$, $\beta_0$, $L$ and $K$. 
In this paper we report the results of simulations for parameters
$\beta_0 = 1$, $\gamma=5/3$, $L=1$, and $K=0.05$. The density of 
the halo at $R=r_0$ is taken to be $\rho_{\rm halo}/\rho_0=10^{-3}$. 
The unit field strength is $B_0 = \rho_0^{1/2}v_{K0}$.

We solve the ideal MHD equations in a cylindrical coordinate system
by using a modified Lax-Wendroff scheme with artificial viscosity. 
We simulated only the upper half space ($z \geq 0$) and assumed that
at the equatorial plane, $\rho$, $v_r$, $v_{\phi}$, $B_r$, 
$B_{\phi}$, and $P$ are symmetric and $v_z$ and $B_z$ are 
antisymmetric. The outer boundaries at $r=r_{\rm max}$ and at 
$z=z_{\rm max}$ are free boundaries where waves can transmit. 
The singularity at $R=0$ is treated by softening the gravitational
potential near the gravitating center ($R<0.2 r_0$). 
The number of grid points is $(N_r, N_{\phi}, N_z) 
= (200, 64, 240)$. To initiate non-axisymmetric evolution, 
small amplitude, random perturbations are imposed at $t=0$ 
for azimuthal velocity. 


\section{NUMERICAL RESULTS}
Figure 1a shows the initial condition. Color scale denotes the density
distribution and red curves depict magnetic field lines. Figure 1b
shows density distribution and velocity vectors at $t=6.2t_0$ where 
$t_0$ is the orbit time $t_0=2\pi r_0/v_{K0}$. As the matter which 
lost angular momentum accretes to the central region, the MHD torus
becomes disk-like. Velocity vectors indicate that matter flows out
from the disk.  

After the non-axisymmetric Balbus \& Hawley instability grows, the
inner region of the torus becomes turbulent. The turbulent motions
tangle magnetic field lines in small scale and create numerous current
sheets (or current filaments). Figure 1c shows the magnetic field
lines projected onto the equatorial plane and the density distribution
at $z=0$. Note that Figure 1c shows only the inner region where $-2.5
\leq x / r_0 \leq 2.5$ and $-2.5 \leq y /r_0 \leq 2.5$. In large
scales, magnetic field lines and density distribution show low
azimuthal wave number spiral structure. Figure 1d shows that magnetic
loop structures similar to those in the solar corona are created. The
yellow surfaces show strongly magnetized regions where
$|\mbox{\boldmath $B$}|=0.1B_0$ and red curves show magnetic field
lines at $t = 7.5 t_0$. The magnetic loops buoyantly rise from the
disk due to the Parker instability. Numerical results indicate that
small loops which newly appeared above the photosphere (emerging
magnetic loops) develop into expanding coronal loops. We can observe
magnetic loops elongated in the azimuthal direction and loops twisted
by the rotation of the disk. 

Figure 2 shows the isosurfaces of plasma-$\beta$ at $t=6.2 t_0$. 
Orange surfaces show strongly magnetized regions where $\beta =
0.1$. Yellow and green surfaces show the region where $\beta = 1$ and
$\beta =10$, respectively. Inside the disk, strongly magnetized,
low-$\beta$ filaments are created. As we shall show below, low-$\beta$
region occupies a small fraction of the total volume. Intermittent
magnetic structures (filamentary strongly magnetized regions) similar 
to those in the solar photosphere develop in the disk.   

Figure 3a shows the spatial average of the magnetic energy, 
$\log (\langle B^2/8 \pi \rangle / \langle P \rangle)$ (dashed curve), 
$\log \langle B^2 / ( 8 \pi P_0 ) \rangle $ (dash-dotted curve)
and $\log (\langle -B_r B_{\phi}/4 \pi\rangle / P_0)  $ (solid curve)
averaged in $ 0.7 \leq r/r_0 \leq 1.3$ and $0 \leq z/r_0 \leq
0.3$. This figure shows that magnetic energy in the equatorial region
decreases within a few rotation period due to the buoyant escape of 
magnetic flux, and that averaged plasma $\beta$ oscillates
quasi-periodically around $\beta \sim 5$. The ratio of averaged
Maxwell stress to initial equatorial pressure, $\alpha_{th}$ is
$\alpha_{th} \simeq 0.07$ when $6 t_0 < t < 12 t_0$. Figure 3b shows
the time evolution of the accretion rate  
\begin{equation}
\dot M_{\rm acc}= \int_0^{2 \pi } \int_0^{0.3 r_0} \rho v_r rdz d\varphi
\end{equation}
at $r=0.3 r_0$, and of the outflow rate
\begin{equation}
\dot M_{\rm out}=\int_0^{2 \pi} \int_0^{r_{\rm max}} \rho v_z rdr
d\varphi 
\end{equation}
at $z=3.0 r_0$.
Accretion rate increases and fluctuates around a mean value. Numerical
results also indicate quasi-periodic ejection of the disk material.
Figure 3c shows the Poynting flux $(\mbox{\boldmath $E$} \times
\mbox{\boldmath $B$} / 4 \pi)_z$ which passes through the plane at 
$z=1.0, 2.0, 3.0 r_0$. After a few rotation period, magnetic flux is 
convected from the equatorial region to the disk surface. The volume
filling factor of the region where $\beta \leq 0.3$ is shown in
Figure 3d. Strongly magnetized region where $\beta \leq 0.3$ occupies
about 8\% of the total volume after a few rotation period. Later, the
filling factor decreases to 2\%. After about 10 rotation period,
magnetic flux is regenerated in the disk and filling factor shows the
second peak when the averaged $1 / \langle \beta \rangle$ is
maximum. Following this second peak, Poynting flux at $z = 2 r_0$
increases, which suggest that magnetic flux escapes from the disk.

\section{DISCUSSION}
We showed by 3D global MHD simulations that when a differentially
rotating torus is threaded by equipartition strength toroidal magnetic 
fields, magnetic loops emerging from the disk continue to rise and
form coronal magnetic loops similar to those in the solar corona.
Inside the disk, magnetic turbulence drives dynamo action
which maintains magnetic fields and keeps the disk in a quasi-steady
state with $\beta \sim 5$. We successfully simulated the formation
process of the solar type disk by 3D direct MHD simulations. 

Numerical results indicate that in differentially rotating disks 
magnetic field lines globally show spiral structure with low azimuthal
wave number but locally fluctuating components create numerous current
sheets. We expect that magnetic reconnection taking place in current 
sheets generate hot plasmas which emit hard X-rays. When the
disk is optically thin, such reconnection events may be observed as 
large amplitude sporadic X-ray time variations characteristic of
low-states in black hole candidates (Kawaguchi et al. 1999).

We found that inside the torus, filamentary shaped, locally strongly 
magnetized, low-$\beta$ regions appear. Even when $\beta \sim 5$
in average, low-$\beta$ regions where $\beta \leq 0.3$ occupy 2-10\%
of the total volume. The low-$\beta$ filaments are embedded in 
high-$\beta$ plasma. Such an intermittent structure is common in 
magnetized astrophysical plasmas. Numerical results indicate that
low-$\beta$ filaments are re-generated after they once disappear.

Although we assumed point gravity suitable for accretion disks,
numerical results can qualitatively be applied to galactic gas disks.
Our numerical results suggest that although $\beta \ge 1$ in average,
low-$\beta$ filaments exist in galactic gas disks. Magnetic 
reconnection taking place in such low-$\beta$ regions may 
heat the interstellar gas and create hot, X-ray emitting plasmas.
Tanuma et al. (1999) proposed a model that magnetic reconnection
in strongly magnetized ($\sim 30\mu G$) regions in our Galaxy creates
hot plasma which emit 7KeV component of Galactic Ridge X-ray Emission
(GRXE). The low-$\beta$ filaments can also confine the hot plasma and
prevent it from escaping from the Galaxy.

\acknowledgments

We thank Drs. K. Shibata, T. Tajima, S. Mineshige, T. Kawaguchi and 
K. Makishima for discussion. Numerical computations were carried out
by VPP300/16R at NAOJ. This work is supported in part by the Grant of 
Japan Science and Technology Corporation and the Grant-in-Aid of the
Ministry of Education, Science, Sports, and Culture, Japan(07640348,
 10147105).

\clearpage

\figcaption[fig1.eps]{The density distribution and magnetic field
lines in a model starting from equipartition strength ($\beta_0
= 1$) toroidal field. (a) The initial state. Color scale 
shows the density distribution and red curves depict magnetic field 
lines. (b) Density distribution at $t=6.2t_0$ and velocity vectors in
$xz$-plane. (c) Density distribution at $z=0$ (color scale) and magnetic
field lines projected onto the equatorial plane ($ -2.5 \leq x/r_0
\leq 2.5, -2.5 \leq y/r_0 \leq 2.5$). (d) Isosurfaces of magnetic field 
strength $|B/B_0| = 0.1$ (yellow) at $t = 7.5 t_0$. The blue surface
shows the slice in $xy$-plane at $z=1.65r_0$. The red curves show
magnetic field lines. \label{fig1}}   

\figcaption[fig2.eps]{ The isosurface of plasma $\beta$ ($ -2.5 \leq
r/r_0 \leq 2.5$, $-2.5 \leq z/r_0 \leq 2.5$) at $t=6.2t_0$. The orange
surfaces show the strongly magnetized region where $\beta=0.1$. The
yellow and green surfaces show the isosurface $\beta=1$ and
$\beta=10$, respectively.  \label{fig2}}     

\figcaption[fig3.eps]{(a) Time development of the mean magnetic energy 
 averaged in the region where $0.7 \leq r/r_0 \leq 1.3$ and $ 0 \leq
z/r_0 \leq 0.3$. The dashed curve, dash-dotted curve and solid curve
show $\log (\langle B^2/8 \pi \rangle / \langle P \rangle)$,
$\log \langle B^2/(8 \pi P_0)\rangle$, and $\log \langle -B_r
B_{\phi}/( 4 \pi P_0) \rangle$, respectively. (b) Time development of
accretion rate at $r=0.3r_0$ and outflow rate at $z=3.0r_0$. (c)
Poynting flux which go through the plane $z=1.0, 2.0, 3.0r_0$. (d)
Volume filling factor of the region where $\beta < 0.3$ in $0.7 \leq
r/r_0 \leq 1.3$ and $0 \leq z/r_0 \leq 0.7$.  \label{fig3}} 


\clearpage

\begin{thebibliography}{}
\bibitem[Balbus and Hawley 1991]{bal91} Balbus, S. A.,
    \& Hawley, J. F.  1991, \apj, 376, 214
\bibitem[Brandenburg et al.,\ 1995]{bra95} Brandenburg, A.,
    Nordlund, A., Stein, R., \& Torkelsson, U. 1995, \apj,
    446, 741
\bibitem[Foglizzo and Tagger 1995]{fog95} Foglizzo, T., \&
    Tagger, M. 1995, \aap, 301, 293
\bibitem[Galeev et al., \ 1979]{gal79} Galeev, A. A., Rosner, R., \& 
    Vaiana, G. S. 1979, \apj, 229, 318 
\bibitem[Hawley et al.,\ 1995]{haw95} Hawley, J. F., Gammie, C. F.,
    \& Balbus, S. A. 1995, \apj, 440, 742
\bibitem[Hawley 1999]{haw99} Hawley, J. F. 1999, \apj, submitted
\bibitem[Kato et al., \ 1998]{kat98} Kato, S., Fukue, J., \& 
    Mineshige, S. 1998, Black-Hole Accretion Disks (Kyoto Univ. Press) 
\bibitem[Kawaguchi et al., \ 1999]{kaw99} Kawaguchi, T., Mineshige, 
    S., Machida, M., Matsumoto, R., \& Shibata, K. 1999, 
    submitted to \pasj
\bibitem[Matsumoto et al., \ 1988]{mat88} Matsumoto, R., 
    Horiuchi, T., Shibata, K., \& Hanawa, T. 1988, \pasj, 40, 171 
\bibitem[Matsumoto and Tajima 1995]{mat95} Matsumoto, R., 
    \& Tajima, T. 1995, \apj, 445, 767
\bibitem[Matsumoto 1999]{mat99} Matsumoto, R. 1999, Numerical 
    Astrophysics, ed. Miyama, S., Tomisaka, K., and Hanawa, T.,
    (Kluwer Academic Publishers), p.195
\bibitem[Okada et al.,\ 1989]{oka99} Okada, R., Fukue, J.,
    \& Matsumoto, R. 1989, \pasj, 41, 133
\bibitem[Parker 1966]{par66} Parker, E. N. 1966, \apj, 145, 811
\bibitem[Shibata et al. \ 1990]{shi90} Shibata, K., Tajima, T.,
    \& Matsumoto, R. 1990 \apj, 350, 295 
\bibitem[Stone et al.,\ 1996]{sto96} Stone, J. M., Hawley, J. F.,
    Gammie, C. F., \&  Balbus, S. A. 1996 \apj, 463, 656
\bibitem[Tanuma et al., \ 1998]{tan98} Tamura, S., Yokoyama, T,
    Kudoh, T., Matsumoto, R., Shibata, K., \& Makishima, K.,
    1999, \pasj, 51, 161
\end{thebibliography}
\end{document}